\documentclass[aps,preprint,showpacs,superscriptaddress]{revtex4}
\usepackage{psfrag,graphicx}
\begin{document}

\newcommand{\bra}[1]{\langle #1 |}
\newcommand{\ket}[1]{| #1 \rangle}
\newcommand{\bk}[2]{\bra{#1} #2 \rangle}
\newcommand{\bok}[3]{\bra{#1} #2 \ket{#3}}
\newcommand{\kb}[2]{\ket{#1} \bra{#2}}
\newcommand{\kk}[1]{\kb{#1}{#1}}
\newcommand{\norm}[1]{\{#1\}}
\newcommand{\Half}{\frac{1}{2}}

\title{Calculation of single-beam two-photon absorption transition rate
 of rare-earth ions using effective operator  and diagrammatic representation
}

\author{Chang-Kui Duan}
\affiliation{
Institute of Applied Physics and College of Electronic Engineering,
Chongqing University of Posts and Telecommunications, Chongqing 400065, 
China.
}
\affiliation{
Department of Physics and Astronomy, University of Canterbury, Christchurch,
  New Zealand}
\author{Michael F. Reid}
\affiliation{
Department of Physics and Astronomy, University of Canterbury, Christchurch,
  New Zealand}
\author{Gang Ruan}
\affiliation{
Institute of Applied Physics and College of Electronic Engineering,
Chongqing University of Posts and Telecommunications, Chongqing 400065, 
China.
}
\author{Hui-Ning Dong}
\affiliation{
Institute of Applied Physics and College of Electronic Engineering,
Chongqing University of Posts and Telecommunications, Chongqing 400065, 
China.
}
\date{\today}

\begin{abstract}
Effective operators needed in single-beam two-photon transition calculations have
been represented with modified Goldstone diagrams similar to the type suggested
by Duan and co-workers [J. Chem. Phys. 121, 5071 (2004) ].  The rules to
 evaluate these diagrams  are different from those for effective Hamiltonian
 and one-photon transition 
operators.   It is verified that the perturbation terms considered contain only connected
diagrams and the evaluation rules are simplified and given explicitly.

\pacs{71.20.Eh, 32.70.Cs, 71.27.+a, 32.80.Wr, 04.25.Wx}
\end{abstract}

\maketitle

\section{Introduction}

Two-photon laser spectroscopy has been used as an important complementary 
technique in studying optical materials containing lanthanide ions.  A recent
short review can be found in Ref. \onlinecite{DuaR2002}.   Two-photon excitation  in a
conventional microscope studying biological samples  provides great advantages,
including low photo-damage and depth discrimination,  reduced photobleaching, 
good separation of the excitation light from
 the fluorescence emission. \cite{Cah2002} Two-photon
lasers could  give rise to entangled twin beams which offer potential for novel
future applications.\cite{Cum1999}  A theoretical challenge in
the field of two-photon  spectroscopy is  prediction of two-photon
 transition rates.\cite{Dow1989,DuaR2002,Bur2003}

Our work  focuses on the calculation of single-beam two-photon absorption
rates,  particular for optical materials with lanthanides ions. In a former
paper,\cite{Dua2004c}  we have presented the effective operator 
method and given a perturbation expansion.   In this paper we show that the 
perturbation expansion can be represented with a kind of diagrams, which can be
calculated using one and two-particle matrix elements of interactions and 
single-particle energies. In Section II,
we show how to derive the diagrams from the perturbation expansion;  Section III
presents those diagrams with only valence internal lines and unusual energy
denominators; In Section IV we show the cancellation of disconnected
diagrams. 

\section{Construction of diagrams from perturbation expansions}
The calculation of two-photon transition rates has been reduced to
 the calculation of an effective transition operator\cite{Dua2004c}. The 
square modulus of the matrix element of the effective operator between
two eigenstates of the effective Hamiltonian (usually approximated with
semi-classical phenomenological Hamiltonian) is the two-photon
transition line strength that can be compared to measurements.
 In Ref. \cite{Dua2004c}, the effective operator 
is partitioned into two terms. The first term is calculated with
 ``direct method''\cite{Dua2004c,Bur2003},
and the second term is calculated using many-body perturbation theory.
It has been shown Ref.\cite{Bur2003} that this method contains more general
 contributions than previous methods.
 Here we start from that expansion for the second term 
and consider the usual case where the model space $P_{I0}$ and $P_{F0}$ 
are the same complete model space, denoted as  $P_0$.  By 
``complete model space'' we mean that 
the model space contains all the bases that can be formed by distributing the 
valence electrons among the valence single particle states.\cite{DuaR2001,Lin1974}
Denoting the space orthogonal to $P_0$ by $Q_0$,  the perturbative expansion
 is simplified into the following form
\begin{eqnarray}
\label{first}
&&\bok{f} {T^2_{\rm eff,0}} {i} =
    \frac {\bok {f}{O}{k}\bok {k}O{i}}
          {(E_{f0} + E_{i0} )/2 - E_{k0}}\\
\label{second}
&&\bok f {T^2_{\rm eff,1}} i =
    \frac{\bok f V {k_1} \bok {k_1} O {k_2} \bok {k_2} O i}
       {(E_{f0}-E_{k_10})[\frac{E_{f0}+E_{i0}}{2}-E_{k_20}]}
\\
\label{third}
 &&
~~~+  \frac{\bok f O {k_1} \bok {k_1} O {k_2} \bok {k_2} V i}
       {[\frac{E_{f0}+E_{i0}}{2}-E_{k_10}](E_{i0}-E_{k_20}) } 
  \\
\label{fourth}
&&
~~~ + \frac{\bok f O l \bok l V k \bok k O i}
       {[\frac{E_{f0}+E_{i0}}{2}-E_{k0}](E_{k0}-E_{l0})}
\\
\label{fifth}
&&
 ~~~ +\frac{\bok f O k \bok k V l \bok l O i}
       {(E_{k0}-E_{l0})[\frac{E_{f0}+E_{i0}}{2}-E_{k0}]}
 \\
\label{sixth}
&&
~~~+ \frac{\bok f O {k_1} \bok {k_1} V {k_2} \bok {k_2} O {i}}
       {[\frac{E_{f0}+E_{i0}}{2}-E_{k_10}][\frac{E_{f0}+E_{i0}}{2}-E_{k_20}]}
\\
\label{seventh}
&&
~~~-\frac{1}{2} 
  \frac{\bok f V {l} \bok {l} O {k} \bok {k} O i}
       {[\frac{E_{f0}+E_{i0}}{2}-E_{k0}][\frac{E_{l0}+E_{i0}}{2}-E_{k0}]}
\\
&&~~~ -\frac{1}{2} 
  \frac{ \bok {f} O {k} \bok {k} O {l} \bok{l} V {i}}
       {[\frac{E_{f0}+E_{l0}}{2}-E_{k0}][\frac{E_{f0}+E_{i0}}{2}-E_{k0}]},
\label{last}
\end{eqnarray} 
where $T^2_{{\rm eff},i}$ is the part of effective operator zeroth ($i=0$) 
 and first oder ($i=1$)  in $V$,
$k$, $k_1$ and $k_2$   sum over  many-particle orthonormal 
eigenstates of $H_0$  in  $Q_0$ and $l$ sums over eigenstates of
$H_0$ in $P_0$. $f$ and $i$ are orthonormal eigenstates of $H_0$ in $P_0$.

The perturbation $V$ contains both a one-body operator, denoted
as $V_1$ and a two-body operator, denoted as $V_2$.  Usually the
operator $O$ is one-body operator.   Using algebraic representation 
and corresponding diagrammatic representation given in Ref.\onlinecite{Dua2004b},
Eq. 1 and Eq.4, $O$,  $V_1$ and $V_2$ can be represented by 4,  4 and 9 diagrams
respectively.  Compositions of two $O$'s and one $V_1$ or $V_2$,  as given
in Eq.\ref{second}--\ref{last},   with all possible contractions and all
possible relative positions, result  into hundreds of diagrams. This prompts us
to adapt the modified diagrams given in Ref.\onlinecite{Dua2004b} to combine the
set of diagrams with the same set of vertices, line directions and contractions
but different relative position and internal orbit types into one diagram.  

\subsection{Adapted connected diagrams with vertices $V_1$ and two $O$}

There are altogether 3 ingoing and 3 outgoing lines, with at least 2 pairs
of them contracted to form connected diagrams. Therefore the connected 
diagram can only be zero-body (a constant) or one-body diagrams. A constant
will not contribute to any transition, which is neglected throughout of this
paper.

\begin{figure}
\vskip 2cm
\centerline{\includegraphics[width=10cm]
                                             {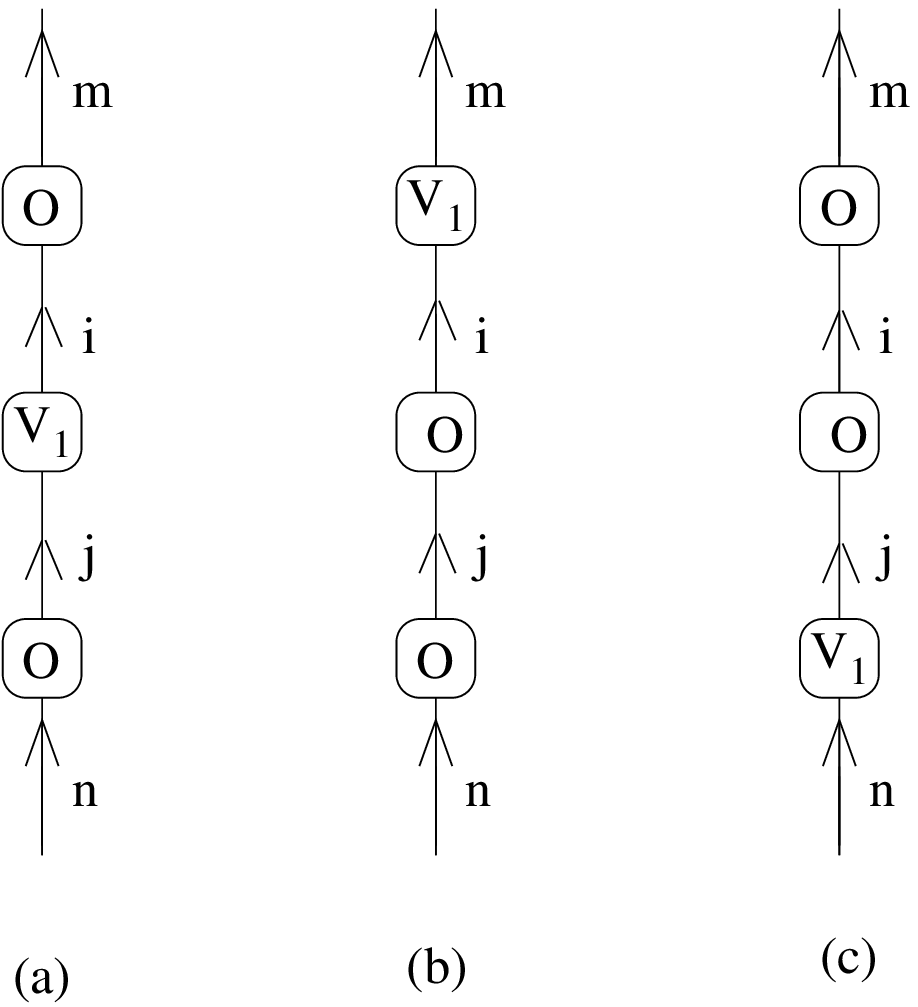}
                         }
\vskip 2cm
\caption{
\label{v1oo}
Modified one-body diagrams first order in $V_1$ for one-beam two-photon
 transition moment. Note that: 1) the internal lines can be both particle
 lines and core lines, and the relative altitudes of two $O$'s and $V_1$
change accordingly; 2) contributions of (b) and (c) to the effective operator are complex
conjugation of each other.
}
\end{figure}

There are three modified  one-body diagrams as shown in Fig. \ref{v1oo}. 
Similar to the the case for effective one-photon transition operators in Fig. 4 of 
Ref. \onlinecite{Dua2004b},  each diagram actually represents  contributions
 from several different terms in Eq.\ref{second}-Eq.\ref{last}. For example
Fig. \ref{v1oo}(a), with different internal line types, includes contributions
from Eq.\ref{second}, Eq.\ref{third} and Eq.\ref{sixth}. Note that the energy
denominator rules will be slightly different from those for effective 
one-photon transition  operators in Ref.\onlinecite{Dua2004b}.

\subsection{Adapted connected diagram with vertices $V_2$ and two $O$}
There are altogether 4 ingoing and 4 outgoing lines for for one $V_2$ and two 
$O$.  Therefore connected diagrams can be zero-body, one-body and two-body.

\begin{figure}
\vskip 2cm
\centerline{\includegraphics[width=12cm]
                                            {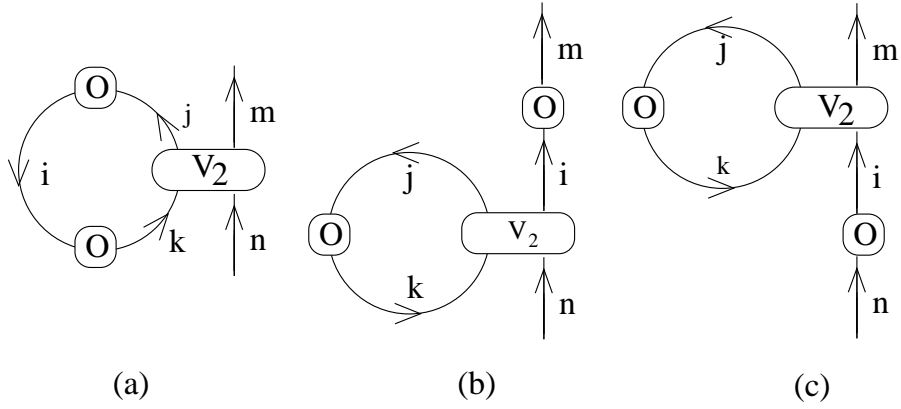}
}
\vskip 2cm
\caption{\label{v2oo-one-body}
Modified one-body diagrams first order in $V_2$ for one-beam two-photon
 transition moment. Note that: 1) the internal lines can be both particle
 lines and core lines, and the relative altitudes of two $O$'s and $V_1$
change accordingly; 2) there are natural constrains over several internal
 lines forming a close loop, since particle lines go upward and core lines
go downward,  for example, in (b), $\{j,k\}$ must be one
particle line and one hole line;  3) contributions of (b) and (c)
to the effective operator are complex conjugation of each other.}
\end{figure}

\begin{figure}
\vskip 2cm
\centerline{\includegraphics[width=12cm]
                             {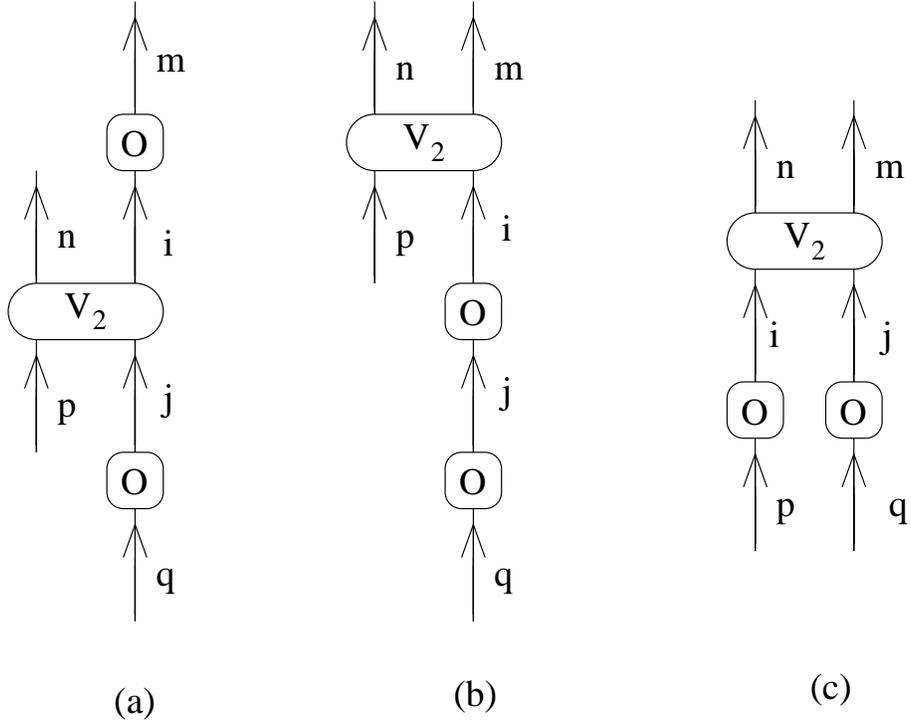}
}
\vskip 2cm
\caption{\label{v2oo-two-body}Modified two-body diagrams first order in $V_2$ for one-beam two-
photon
 transition moment. Note that: 1) the internal lines can be both particle
 lines and core lines, and the relative altitudes of two $O$'s and $V_1$
change accordingly; 2) two diagrams with contributions complex conjugate to  these of (b) and (
c)
are not shown here.
}
\end{figure}

There are also three modified one-body diagrams as shown in Fig. \ref{v2oo-one-body}
, and five modified two-body diagrams as shown in Fig. \ref{v2oo-two-body},
where two diagrams complex conjugate to Fig.\ref{v2oo-two-body}(b) and (c) are
not given explicitly.

Similar to  connected diagrams in Fig.\ref{v1oo},  Fig.\ref{v2oo-one-body}(a),
 includes contributions from Eq. \ref{second}, \ref{third},
\ref{sixth}. These contributions form  altogether 6 different normal diagrams.
 Note that the energy denominator
rules for contributions from different terms are slightly different. 

\subsection{Rules to evaluate these diagrams}

Remembering that each modified diagram actually represents a set of diagrams
with different ordering of vertices and different types of
 internal lines,  it is not difficult to obtain the
analytic expressions for these diagrams.  For each diagram
 with fixed vertex ordering, the expression
can be calculated by the same rules 1)--7) given in Sec. II, Ref. 
\cite{Dua2004a}, except that the rule to
calculate energy denominators in rule 6) should be modified as follows:
 Denoting  the net outflow energy of a loop
enclosing vertices $1,~2,\cdots, ~i$ as
\begin{equation}
 E_{\rm noe}(1,~2,\cdots,~i) = E_{\rm noe}(1)+E_{\rm noe}(2)+\cdots +E_{\rm noe}(i),
\end{equation}
and supposing the number of $O$  in the loop is $i_{O} ~(i_{O}=0,~1,~2)$ 
 and the number of vertices for the diagram is $n$, the energy
 denominator can be written as
\begin{eqnarray}
D = \prod\limits_{i=1}^{n-1} \left (\frac{i_{O}}{2}
E_{\rm noe}(1,~2,\cdots,~n)-E_{\rm noe}(1,~2,\cdots,~i)\right ).
\end{eqnarray}

\section{Diagrams with vertex operators acting only on model spaces $P_0$}

Eq. \ref{fourth}, \ref{fifth} \ref{seventh} and \ref{last} give diagrams with some vertices,
whose bra states and ket states are in model spaces $P_0$.
As an example,  Eq. \ref{fourth}   have the same
operator  sequence $OVO$ as that of Eq. \ref{sixth}, but  different denominators. 
 They will give diagrams with different evaluation rules, similar to but not exactly
the same as those terms in effective Hamiltonian that give  folded 
diagrams.\cite{Bra1967,Lin1985,KuoO1990}
  Instead of introducing new folded diagrams,
we combine these contributions to the diagrams given above.  The rules to evaluate 
these contributions differ from standard rules in an overall factor $-1/2$ for
diagrams from Eq.\ref{seventh} and Eq.\ref{last}, and in energy denominators
by using rules given in Eq.\ref{fourth}, Eq.\ref{fifth}, Eq.\ref{seventh} and Eq.\ref{last}.

\section{Cancellation of disconnected diagrams}
For modified disconnected diagrams with one $V$ vertex and two $O$ vertices,  
there must be a connected part with only one vertex, which is either $V$ or $O$.
    
 Case A: $V$ is disconnected from other parts. In this case both the ingoing and the
 outgoing lines of $V$ are valence lines.  Such a diagram includes contributions
from   Eq.\ref{sixth}--\ref{last} with the same matrix elements but different
coefficients due to different energy denominators.  Suppose the net-outgoing-energies
from the first $O$ and the second $O$ vertices are $\Delta E_1$ and $\Delta E_2$ 
respectively, and that from the $V$ is  $\Delta E_3$.  The three different  coefficients
for the three contributions from  Eq.\ref{sixth}--\ref{last}, respectively, are as
follows
\begin{widetext}
\begin{eqnarray}
&&\frac{1}{(\frac{[\Delta E_1 + \Delta E_3  +\Delta E_2)+0}{2}-(\Delta E_3 +\Delta E_2)]
                 [\frac{(\Delta E_1 + \Delta E_3  +\Delta E_2)+0}{2}-\Delta E_2]}\\
&+&(-\frac{1}{2})\frac{1}{[\frac{(\Delta E_3 + \Delta E_1  +\Delta E_2)+0}{2}-\Delta E_2]
                                                  [\frac{(\Delta E_1 + \Delta E_2)+0}{2}-\Delta E_2]}\\
&+&(-\frac{1}{2}) \frac{1}{[(\frac{(\Delta E_1 +\Delta E_2+\Delta E_3)+0}{2}-(\Delta E_2 + 
                                                                                                                              \Delta E_3)]
                                                  [\frac{(\Delta E_1 +\Delta E_2+\Delta E_3)+\Delta E_3}{2}
                                                                                                   -(\Delta E_2 + \Delta E_3)]}\\
&=&0
\end{eqnarray}
\end{widetext}
Case B: one $O$ is disconnected from other parts.  Such a diagram includes contributions
from Eq. \ref{second}--\ref{fifth}. It can be verified in the same way as 
above that all these contributions cancel.

\section{Conclusion}
Diagrams for  two-photon transitions rates have been constructed and 
have been shown to be connected.  Simplified one-photon diagrams given
 in J. Chem. Phys.  121, 5071 (2004) have been adapted to two-photon cases. 
 The algebraic expressions suitable for a general two-photon calculation 
can be given from these diagrams.
\section*{Acknowledgment}

CKD acknowledges support of this work by the Nation Foundation
of Nature Science (China), Grant No. 10404040 and No. 10274079.

\bibliography{FP-78}


\section*{ Figures}

\end{document}